\documentclass[aps,pre,twocolumn,float]{revtex4}
\usepackage{amsmath,bm,epsfig}
\usepackage{graphicx}

\newcommand{\B}[1]{{\bm{#1}}}%% Bold Roman & Greek Lower & Upper Case
\def\Xint#1{\mathchoice
   {\XXint\displaystyle\textstyle{#1}}%
   {\XXint\textstyle\scriptstyle{#1}}%
   {\XXint\scriptstyle\scriptscriptstyle{#1}}%
   {\XXint\scriptscriptstyle\scriptscriptstyle{#1}}%
   \!\int}
\def\XXint#1#2#3{{\setbox0=\hbox{$#1{#2#3}{\int}$}
     \vcenter{\hbox{$#2#3$}}\kern-.5\wd0}}
\def\Cint{\Xint C}

\begin{document}
\title{Stress Intensity Factor of Mode III Cracks in Thin Sheets.}
\author{Yossi Cohen and Itamar Procaccia}
\affiliation{Department of Chemical Physics, The Weizmann
Institute of Science, Rehovot 76100, Israel  }
\date{\today}
\begin{abstract}
The stress field at the tip of a crack of a thin plate of elastic
material that is broken due to a mode III shear tearing has a
universal form with a non-universal amplitude, known as the stress
intensity factor, which depends on the crack length and the
boundary conditions. We present in this paper exact analytic
results for this stress intensity factor, thus enriching the
small number of exact results that can be obtained within Linear
Elastic Fracture Mechanics (LEFM).
\end{abstract}
\maketitle
\section{Introduction}
Linear Elastic Fracture Mechanics (LEFM) \cite{98F} deals with the dynamics
of cracks in elastic materials subject to the assumption that
linear elasticity if applicable everywhere in the material except
for a very small region around the crack tip known as the process
zone. While recent experiments and theories indicate that there
exist examples where nonlinear elasticity may be important for a
full characterization of the stress field before a dynamical
crack, some exact results of LEFM continue to have a very
important role in describing and understanding the dynamics of
cracks in a variety of fracture contexts.  Indeed, one of the
most celebrated results of LEFM is the analytic description of
the universal stress distribution in front of a crack. This
analytic result is particularly simple when the crack is
developing under one of the three fundamental shearing modes that
a material can suffer. The simplest form results from mode III
cracking which is obtained under an out-of-plane shear tearing.
In a cylindrical coordinate system $r$, $\theta$, $z$, where $r=0$
coincide with the crack tip and $\theta=\pm \pi$ with the crack
faces, the stress in front of a crack takes on the form
\begin{equation}
\sigma_{\theta z}=\frac{K_{III}}{\sqrt{2\pi
r}}\cos\left(\frac{\theta}{2}\right) \ , \label{regular}
\end{equation}
where $K_{III}$ is the so-called ``Stress Intensity Factor". The
results (\ref{regular}) underlines the universality of the angular
dependence and of the square-root singularity of the stress at
the crack tip \cite{74GS,99B}. All the non-universal aspects of the stress
distribution are collected in the Stress Intensity Factor which
depends on everything, including the crack length, the boundary
conditions and the history of the loads that drive the crack evolution. The value of the stress intensity factor
determines whether the crack propagates or not;  Irwin \cite{57I} established a relationship between the energy release rate and the stress intensity factor
for a crack growth. A crack propagates
as long as the energy release rate into the crack tip balance the
dissipation involved in the crack propagation, i.e. the Griffith
criterion \cite{24G,98F}.

Solving for mode III is relatively easy since the fundamental equation to be dealt with is the
Laplace equation \cite{61BC} rather than the bi-Lapalace equation which is the relevant one for modes I and II. On the other hand
in 3-dimensional materials mode III is unstable and attempts to break in this mode quickly turn to cracks developing under effective
modes I or II. This is not the case for thin plates where mode III can be sustained in experiments
as had been demonstrated recently in Ref. \cite{10BBA}. The theory of thin plates is however more cumbersome, using again the bi-Laplace operator even for mode III cracks. Nevertheless the problem can be treated analytically \cite{93HZ,98ZHCH,10CP} with the result that the leading singularity at the crack tip is higher, going like $r^{-3/2}$. In this case, the shear force close to the crack tip can be written as
\begin{equation}
Q_\theta=\int_{-\frac{h}{2}}^{\frac{h}{2}}\sigma_{\theta z}dz=\frac{\tilde K_{III}}{(\pi r)^{3/2}}\cos\left(\frac{\theta}{2}\right),
\end{equation}
we assign tilde on $K_{III}$ to distinguish it from the known stress intensity factor for this mode. $h$ is the thickness of the plate.

We may mention that the Kirchhoff plate theory predicts the stresses very accurately for small deflection and at some distance from the crack tip $(r/h>1)$ (cf. \cite{05ZV} and references therein). However, for large deflection or near the crack tip $(r/h<0.1)$, the singularity of the stress field weakens \cite{05ZV,10CP}, and therefore it may diverge from the actual results.

The aim of the present paper is to present further analytic progress that culminates in the calculation of the stress intensity factor
for cracks developing in thin sheets.
The structure of the paper is as follows: in Sect. \ref{setup} we set up the problem for mathematical analysis. In Sect. \ref{bc}
we study the properties of the analytic functions that need to be solved for and the boundary conditions. In Sect. \ref{conformal}
we introduce the conformal map from the strip to the upper half plane which is used to advantage in simplifying the solution
of the problem. Sect. \ref{SIF} wraps up the problem and presents the analytic result for the Stress Intensity Factor.

\section{Setting up the problem}
\label{setup}

\begin{figure}
  \includegraphics[scale=0.3]{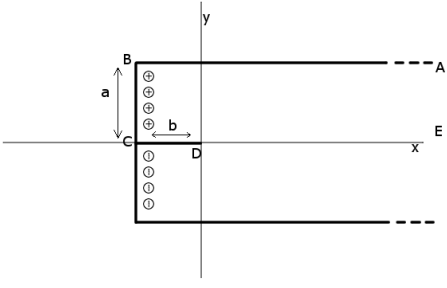}
  \caption{The mode III configuration. A semi-infinite strip of width $2a$, and a cut (C-D) of length $b$ in the middle of the strip. The upper left side of the strip (B-C) is subjected  to an anti-plane positive load, and the lower left side to an identical but negative load.}\label{sheet}
\end{figure}

Consider a semi-infinite strip of a thin sheet with a cut in the in the middle of the left-most end of the strip, along the symmetry axis.
The left boundary is  subjected to anti-symmetric shear load $f_b$, i.e. mode III configuration, cf. Fig. \ref{sheet}.
For small deflection, we can neglect the in-plane deformation of the middle surface. According to the Kirchhoff
plate theory \cite{40T}, the deflection $w$ in the $z$-axis for pure bending becomes
\begin{equation}
\triangle^2 w=0 \ ,
\label{bh}
\end{equation}
where the operator `delta' stands for the Laplace operator $\B \nabla\cdot \B \nabla$;
the bending moments and the twisting moment are
\begin{eqnarray}
M_x&=&\int_{-\frac{h}{2}}^{\frac{h}{2}}\sigma_{xx} zdz=-D\left(\frac{\partial^2 w}{\partial x^2} + \nu\frac{\partial^2 w}{\partial y^2}\right), \nonumber \\
M_y&=& \int_{-\frac{h}{2}}^{\frac{h}{2}}\sigma_{yy}zdz=-D\left(\frac{\partial^2 w}{\partial y^2} + \nu\frac{\partial^2 w}{\partial x^2}\right), \nonumber \\
M_{xy}&=& \int_{-\frac{h}{2}}^{\frac{h}{2}}\sigma_{xy}zdz=D(1-\nu)\frac{\partial^2 w}{\partial x\partial y},
\label{moments}
\end{eqnarray}
and the anti-plane shear stress
\begin{eqnarray}
Q_x&=&\int_{-\frac{h}{2}}^{\frac{h}{2}}\sigma_{xz}dz = -D\frac{\partial}{\partial x}(\triangle w), \nonumber \\
Q_y&=&\int_{-\frac{h}{2}}^{\frac{h}{2}}\sigma_{yz}dz = -D\frac{\partial}{\partial y}(\triangle w).
\end{eqnarray}
where the sheet is on the $xy$ plane, and the bending occurs along the $z$ direction. $\sigma_{ij}$ is a component of the stress tensor, $D=Eh^3/12(1-\nu^2)$ is the so-called \emph{the flexural rigidity} of a plate, $h$ is the thickness of the plate, and $E$, $\nu$ are the Young's modulus and Poisson's ratio, respectively.

Introduce now the complex notation $z=x+iy$, $\bar z =x-iy$, such that
\begin{eqnarray}
\frac{\partial}{\partial z} &=&\frac{1}{2} (\partial_x -i\partial_y)\equiv\partial_z,\\
\frac{\partial}{\partial \bar z} &=&\frac{1}{2} (\partial_x +i\partial_y)\equiv\partial_{\bar z}.
\end{eqnarray}

Since $w$ is a solution of the bi-Laplace equation, $w$ can be expressed as \cite{53M}
\begin{equation}
w(z)=\Re [\bar z\varphi(z) + \chi(z)],
\end{equation}
where $\Re$ denotes ``the real part", and $\varphi(z), \chi(z)$ are both analytic functions, accept maybe on the boundary.

The complex first derivative becomes
\begin{eqnarray}
\partial_x w +i\partial_y w&=&2\partial_{\bar z} w \nonumber \\
 &=& z \overline {\varphi'(z)} + \overline{ \chi'(z)} + \varphi(z).
\end{eqnarray}

The bending moments and the twisting moment are
\begin{eqnarray}
M_x\!\!&=&\!\!-D\left[(1\!-\nu)\Re\left(\bar z\varphi''+\chi''\right)+2(1+\nu)\Re\left(\varphi'\right)\right], \\
M_y\!\!&=&\!\!-D\left[\!-\!(1\!-\nu)\Re\left(\bar z\varphi''\!+\!\chi''\right)+2(1+\nu)\Re\left(\varphi'\right)\right],\\
M_{xy}\!\!&=&\!\!D(1-\nu)\Im\left(\bar z\varphi''+\chi''\right).
\end{eqnarray}

Summing over the two bending moments gives
\begin{eqnarray}
M_x+M_y&=&-D(1+\nu)\triangle w \nonumber \\
&=&-4D(1+\nu)\Re(\varphi').
\label{summ}
\end{eqnarray}
and the complex representation of the shear force becomes
\begin{equation}
Q_x+iQ_y=-4D\overline{\varphi''(z)}
\label{Q}
\end{equation}

Next, the shear force exerted normal to the boundary is
\begin{equation}
Q_n=Qx\frac{dy}{ds}-Qy\frac{dx}{ds}.
\end{equation}
where $n,s$ are the normal and the tangent direction oriented with respect to each other as the axes $x,y$.
Multiply the last equation by $ds$ and using the complex representation in Eq. \eqref{Q} gives
\begin{eqnarray}
Q_nds&=&Q_xdy-Q_ydx \nonumber \\
&=&2Di(\varphi''dz - \overline{\varphi''dz}).
\label{Qn}
\end{eqnarray}

\section{Properties of $\varphi$ and Boundary conditions}
\label{bc}
In the case of mode III symmetry, the load and the deflection are anti-symmetric with respect to the $x$ axis, (cf. Fig. \ref{sheet}.) We consider only one crack lying on part of the $x$ axis (C-D). The upper (B-C) and the lower left edges are displaced with a constant distance from each other, in a direction perpendicular to the strip plane. Then, $w$ is anti-symmetric, but also $\triangle w=4\Re[\varphi'(z)]$ is anti-symmetric. This implies that
\begin{equation}
\Re[\varphi'(z)]=-\Re[\varphi'(\bar z)]
\end{equation}
the left and the right members give
\begin{equation}
\varphi'(z)+\overline{\varphi'(z)}=-\varphi'(\bar z)-\overline{\varphi'(\bar z)}
\end{equation}
which may be arranged to
\begin{equation}
\varphi'(z)+\bar\varphi' (z)=-\overline{\varphi'(z)+\bar\varphi'(z)}.
\label{an}
\end{equation}

Now, if $\varphi(z)$ is analytic on both the upper and the lower half plane, though not necessary on the boundary, the function $\bar \varphi (z)$ is also analytic on the same domain. Thus the left side of Eq. \eqref{an} is analytic in the whole domain (excluding the boundary). Since it equals to minus its complex conjugate, it must be a pure imaginary function. However, according to the Cauchy- Riemann equations, an imaginary analytic function is a constant. Thus,
\begin{equation}
\bar \varphi'(z)=-\varphi'(z) + iC,
\end{equation}
Where $C$ is a real number. Taking the complex conjugate of this expression, followed by the complex conjugation of $z$, gives $C=0$. Thus, the anti-symmetry with respect to the $x$ axis prevails
\begin{equation}
\bar \varphi'(z)=-\varphi'(z).
\label{as}
\end{equation}

Since $\varphi'(z)$ is not necessary analytic on the $x$ axis and on the boundary, it is advantageous to separate notations, $\varphi_+'(z)$ and $\varphi_-'(z)$, for its parts on the upper and lower half-planes,
respectively.

\subsection{Ahead of the crack}
\label{ahead}
In the case of a symmetric body with an anti-symmetric loading, it can be assumed that $w(x,0)=0$ on the symmetry axis ahead of the crack (D-E). Moreover, the anti-symmetry of $w$ dictates that $\triangle w(x>0,0)=0$. From Eqs. \eqref{summ} and \eqref{as}
\begin{eqnarray}
\frac{1}{2}\triangle w_+(x,0)&=&2\Re[\varphi'(z)]=\varphi'(x+i0) + \overline{\varphi'(x+i0)} \nonumber \\
&=&\varphi'(x+i0) + \bar\varphi'(x-i0) \nonumber \\
&=&\varphi'(x+i0) - \varphi'(x-i0)  \nonumber \\
&=&\varphi_+'(x) - \varphi_-'(x)=0
\end{eqnarray}
We used the subscripts $+$ and $-$ for the the limiting values of the functions reaching from the upper and the lower half of the plane, respectively.

\subsection{Crack face}
The crack face (C-D) is assumed to be traction free. Thus, the stress components acting normal to the crack's face become zero. This condition gives
\begin{equation}
My=Mxy=Qy=0, \ for \ C<x<D, y=0
\end{equation}

Integrating Eq. \eqref{Qn} along the boundary from a point $x$ on the crack face to end of the strip, i.e. point $A\rightarrow\infty$, gives
\begin{equation}
\int_{x+i0}^AQ_nds=2Di\int_{x+i0}^A[\varphi''dz-\overline{\varphi''dz}],
\end{equation}
or
\begin{equation}
\int_{x+i0}^A\!\!Q_nds=-2Di\{\varphi'(x+i0)-\overline{\varphi'(x+i0)}-[\varphi'(\infty)-\overline{\varphi'(\infty)}]\}.
\label{int}
\end{equation}

Due to a finite slope at infinity, we can assume that $\lim_{A\to \infty} \varphi'(z=A)=0$. Also, on the crack face and on the upper free edge, $Q_y=0$. Thus, Eq. \eqref{int} becomes
\begin{equation}
\varphi'(x+i0)-\overline{\varphi'(x+i0)}=\frac{i}{2D}\int^{B}_{C}Q_nds.
\end{equation}
and by \eqref{as}, we find
\begin{equation}
\varphi_+'(x) + \varphi_-'(x)=i\beta,
\end{equation}
and
\[
\beta=\frac{f_b}{2D}=\frac{1}{2D}\int^{B}_{C}Q^0_nds.
\]
where $f_b$ is the total applied force on the upper left boundary to create a constant deflection.

\subsection{The loaded edge}
This edge (B-C) is considered as a simply supported edge i.e. $\partial w(-b,y)/\partial y=0$ and $M_x(-b,y)=0$, for $x=-b, \ y>0$ \cite{40T}. These boundary conditions annul also $M_y(-b,y)=0$, and by the same consideration of Sec. (\ref{ahead}) the complex function becomes
\begin{equation}
\varphi_+'(-b+iy) - \varphi_-'(-b-iy)=0
\end{equation}

\subsection{Free upper boundary}
As earlier, a free edge (B-A) feels a zero normal stress. Thus,
\[M_y=M_{xy}=Q_y=0
\]
with the last condition giving
\begin{equation}
\varphi''(x+ia)-\overline{\varphi''(x+ia)}=0, \ for \ -b<x,\ y=a.
\end{equation}

Applying the same integration as in Eq. \eqref{int} for the upper edge gives,
\begin{equation}
\varphi'(x+ia)-\overline{\varphi'(x+ia)}=0
\end{equation}
and by Eq. \eqref{as}, we obtain
\begin{equation}
\varphi_+'(x+ia) + \varphi_-'(x-ia)=0.
\end{equation}

%%%%%%%%%%%%%%%%%%%%%%%%%%%%%%%%%%%%%%%%%%%%%%%%%%%%%%%%%%%%%
\section{Conformal mapping}
\label{conformal}
\begin{figure}
  \includegraphics[scale=0.35]{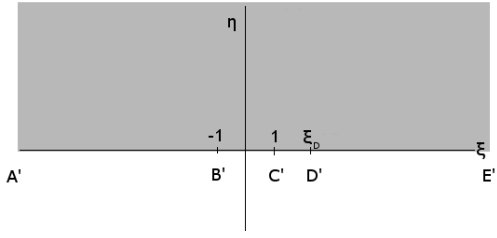}
  \caption{Mapping of the upper strip in Fig \ref{sheet}, onto the upper half of $\zeta$-plane. The inverse function is $\zeta=\cosh\left(\frac{\pi(z+b)}{a}\right)$, and the crack tip is at $\xi_D=\cosh(\pi b/a)$}\label{map}
\end{figure}

The upper half of the strip is mapped onto the upper half of $\zeta$, where $\zeta=\xi+i\eta$. The mapping function is (cf. Fig. \ref{map})
\begin{equation}
z=\Omega(\zeta)=\frac{a}{\pi}\cosh^{-1}(\zeta) - b.
\end{equation}
The analytic function can be written as
\begin{equation}
\varphi'(z)=\varphi'\left(\Omega(\zeta)\right)=F(\zeta).
\label{df}
\end{equation}
and the boundary conditions become
\begin{equation}
\left\{
\begin{array}{lrclr}
F_+(\xi)-F_-(\xi)=0  \ \ &\xi_D<&\xi& &,\eta=0 \\
F_+(\xi)+F_-(\xi)=i\beta \ \  &1<&\xi&<\xi_D&,\eta=0 \\
F_+(\xi)-F_-(\xi)=0 \ \  &-1<&\xi&<1&,\eta=0 \\
F_+(\xi)+F_-(\xi)=0 \ \ & &\xi&<-1&,\eta=0
\end{array}
\right.
\label{BC}
\end{equation}

These four equations constitute a Hilbert problem \cite{53M}. By introducing an auxiliary function $G(\zeta)$, we can write the boundary equations as a single equation. In our case, the function
\begin{equation}
G(\zeta)=(\zeta+1)^{1/2}(\zeta-1)^{1/2}(\zeta-\xi_D)^{1/2}
\end{equation}
with branch cut along $\eta=0$ and $\xi<-1,\ 1<\xi<\xi_D$ with a choice of a branch where $G(\zeta)\rightarrow \zeta^{3/2}$ as $\zeta\rightarrow\infty$.
Then,
\begin{equation}
\frac{G_-(\xi)}{G_+(\xi)}=\left\{
\begin{array}{ll}
+1 \ & for \ -1<\xi<1,\  \xi_D<\xi \\
-1 \ & for \ \ \xi<-1, \ 1<\xi<\xi_D
\end{array}
\right.
\label{G}
\end{equation}
and equations \eqref{BC} can be written as a single equation
\begin{equation}
F_+(\xi) - \frac{G_-(\xi)}{G_+(\xi)}F_-(\xi)=i\beta[H(\xi-1) - H(\xi-\xi_D)]
\label{HP}
\end{equation}
where $H(\xi)$ is Heaviside step function.

Multiplication of the both sides with $G_+(\xi)$ gives
\begin{equation}
G_+(\xi)F_+(\xi) - G_-\!(\xi)F_-\!(\xi)\!=\!iG_+(\xi)\beta[H(\xi-1) - H(\xi-\xi_D)]
\end{equation}

Using the Plemelj's formulae \cite{99B}, we can define an analytic function in both upper and the lower half plane of $\zeta$,
\begin{equation}
G(\zeta)F(\zeta)=\frac{1}{2\pi i}\int^{\xi_D}_1\frac{iG_+(\xi)\beta}{\xi-\zeta}d\xi.
\end{equation}

However, this solution of the Hilbert problem is not complete and we must add an analytic function in the whole plane, since it vanished in the left hand side of Eq. \eqref{HP}. Thus,
\begin{equation}
F(\zeta)=\frac{\beta}{2\pi G(\zeta)}\int^{\xi_D}_1\frac{G_+(\xi)}{\xi-\zeta}d\xi + \frac{P_0(\zeta)}{G(\zeta)}.
\label{F}
\end{equation}
where $P_0(\zeta)$ is analytic in the whole plane and therefore, by Liouville's theorem a polynomial of finite degree.

Since the moments are bounded as $|\zeta|\rightarrow\infty$, the degree of $P(\zeta)$ can not be greater than 1.
In addition, upon the condition in Eq. \eqref{as} the coefficients of the polynomial must be imaginary, i.e. $P_0(\zeta)=i(p_1\zeta + p_0)$, where $p_1$ and $p_0$ are real numbers.

%%%%%%%%%%%%%%%%%%%%%%%%%%%%%%%%%%%%%%%%%%%%%%%%%%%%%%%%%%%%%%%%%%%%%%%%%%%%%%%%%%%%
\section{Shear stress and stress intensity factor}
\label{SIF}
From Eqs. \eqref{Q} and \eqref{df} the shear force in the $\zeta$-plane becomes
\begin{equation}
Q_y=-4D\Im\overline{\varphi''(z)}=4D\Im\frac{F'(\zeta)}{\Omega'(\zeta)},
\label{qydef}
\end{equation}
and $\Omega'(\zeta)=\frac{a}{\pi}\frac{1}{\sqrt{\zeta^2-1}}$ is the map's derivative.

Also, from the relation on Eq. \eqref{as}, the shear force at the point $t>\xi_D$ along the $\xi$-axis becomes
\begin{equation}
Q_y(t)=\frac{-2iD}{\Omega'(t)}\left(F_+'(t)+F_-'(t)\right).
\end{equation}
here, we use the fact that $\Omega'(t)=\overline{\Omega'(t)}$ in this section.

Differentiate the solution of $F(\zeta)$ from the preceding section, Eq. \eqref{F}, gives
\begin{equation}
\begin{aligned}
F'(\zeta)=&-\frac{i\beta G'(\zeta)}{G^2(\zeta)}I(\zeta) + \frac{i\beta}{G(\zeta)} I'(\zeta) \\
&- \frac{i(p_1\zeta+p_0)G'(\zeta)}{G^2(\zeta)} + \frac{ip_1}{G(\zeta)},
\end{aligned}
\end{equation}
where
\[
I(\zeta)=\frac{1}{2\pi i}\int^{\xi_D}_1\frac{G_+(\xi)}{\xi-\zeta}d\xi.
\]
notice that $G_+(\xi)$ is imaginary.

Now, the shear force away from the crack at the symmetry axis is
\begin{equation}
Q_y(t)=\!-\!\frac{4D}{\Omega'(t)G(t)}\bigg[\frac{G'(t)}{G(t)}(\beta I(t)+p_1t+p_0)\!-\!(\beta I'(t)+p_1)\bigg]
\end{equation}
or
\begin{equation}
Q_y(t)=\!-\!\frac{4D\pi}{a\sqrt{t-\xi_D}}\bigg[\frac{G'(t)}{G(t)}(\beta I(t)+p_1t+p_0)-\!(\beta I'(t)+p_1)\bigg]
\label{fQy}
\end{equation}
where
\[
\frac{G'(t)}{G(t)} = \left(\frac{t}{t^2-1}+\frac{1}{2(t-\xi_D)}\right).
\]
$I(t)$ is the Cauchy's principal value,
\[
I(t)=\frac{1}{2\pi i}\Cint^{\xi_D}_1\frac{G_+(\xi)}{\xi-t}d\xi.
\]
and $I'(t)$ is its derivative.

The constants $p_1$ and $p_0$ can be found from the boundary conditions. At the edge $B-C$, $Q_x(t)$ has the same expression as in Eq.\eqref{fQy}. Since we assume a finite applied load at this edge, the singularity of the shear force at $t=\pm1$ disappears if
\begin{equation}
\beta I(t)+p_1t+p_0=0.
\end{equation}
which gives
\begin{eqnarray}
p_1&=&\frac{1}{2}\beta[I(-1)-I(1)],\nonumber \\
p_0&=&-\frac{1}{2}\beta[I(-1)+I(1)].
\label{p10}
\end{eqnarray}

Finally, the stress intensity factor close to the crack tip in thin plates is obtained from
\begin{equation}
\tilde K_{III}=\lim_{x\rightarrow 0}(\pi x)^\frac{3}{2}Q_y(t),
\label{KIIIcm}
\end{equation}
where $x=\Omega(t)$. For $x\rightarrow 0$, we can write
\begin{equation}
x=\frac{a}{\pi\sinh\frac{\pi b}{a}}(t-\xi_D)
\end{equation}
thus, Eq. \eqref{KIIIcm} becomes
\begin{equation}
\tilde K_{III}=\left(\frac{a}{\sinh\frac{\pi b}{a}}\right)^\frac{3}{2}\lim_{t\rightarrow\xi_D}(t-\xi_D)^\frac{3}{2}Q_y(t)
\end{equation}

Applying the expression found in Eq. \eqref{fQy} gives
\begin{equation}
\tilde K_{III}=-\frac{2D\pi\sqrt{a}}{\sinh^\frac{3}{2}(\frac{\pi b}{a})}(\beta I(\xi_D)+p_1\xi_D+p_0).
\end{equation}
using \eqref{p10}
\begin{equation}
\tilde K_{III}=-\frac{D\beta\sqrt{a}(\xi_D^2-1)}{i\sinh^\frac{3}{2}(\frac{\pi b}{a})}\int^{\xi_D}_1\frac{d\xi}{\sqrt{\xi^2-1}\sqrt{\xi-\xi_D}}
\end{equation}
notice that $\xi_D^2-1=\sinh^2(\frac{\pi b}{a})$. Changing the variable of integration to $\lambda=\frac{1}{k}\sqrt{\frac{\xi-1}{\xi+1}}$ with the parameter $k^2=\frac{\xi_D-1}{\xi_D+1}$, one obtains
\begin{equation}
\tilde K_{III}=2D\beta\sqrt{a} \sqrt{\frac{\sinh(\frac{\pi b}{a})}{\xi_D+1}}\int^1_0\frac{d\lambda}{\sqrt{1-\lambda^2}\sqrt{1-k^2\lambda^2}}
\end{equation}
and finally, the stress intensity factor for mode III in thin sheets becomes
\begin{equation}
\tilde K_{III}=f_b\sqrt{a} \sqrt{\frac{\sinh(\frac{\pi b}{a})}{\cosh(\frac{\pi b}{a})+1}}K(k),
\label{KIIIf}
\end{equation}
where $2a$ is the strip width, $b$ is the crack length, $f_b$ is the total load and $K(k)$ is the complete elliptic integral of the first kind.

For a crack larger than the strip's width $(b>>a)$, we can estimate the stress intensity factor in term of the deflection $w_0$ $(\frac{w_0}{b}<<1)$ instead of the load. In this case, $K(k)\sim b/a$ and the stress intensity factor becomes $\tilde K_{III}\sim f_bb/\sqrt{a}$. An estimate of the load from the bending theory gives $f_b\sim \frac{Daw_0}{b^3}$ \cite{86LL}, and we obtain $\tilde K_{III}\sim \frac{D\sqrt{a}w_0}{b^2}$. Hence, in particular, we see that the stress intensity factor decreases as the square of the crack length.

\end{document}